\documentclass[runningheads,a4paper]{llncs}

\usepackage{amssymb}
\usepackage{graphicx}
\usepackage{algorithm}
\usepackage{algorithmic}
\usepackage{hyperref}
\usepackage[numbers]{natbib}

\usepackage{url}
\urldef{\mailsc}\path|Djallel.Bouneffouf}@it-sudparis.eu|

\begin{document}

\mainmatter  % start of an individual contribution

\title{Context-Based Information Retrieval in Risky Environment}

\titlerunning{Lecture Notes in Computer Science: Authors' Instructions}

\author{Djallel Bouneffouf%, Amel Bouzeghoub and Alda Lopes Gançarski   %
\thanks{}
}
\authorrunning{Lecture Notes in Computer Science: Authors' Instructions}

\institute{Department of Computer Science, T\'{e}l\'{e}com SudParis, UMR CNRS Samovar, 91011 Evry Cedex, France,\\
%\mailsa \\
\mailsc\\
%\url{http://www.springer.com/lncs}
}


\toctitle{Lecture Notes in Computer Science}
\tocauthor{Authors' Instructions}
\maketitle

\begin{abstract}
Context-Based Information Retrieval is recently modelled as an exploration/ exploitation trade-off (exr/exp) problem, where the system has to choose between maximizing its expected rewards dealing with its current knowledge (exploitation) and learning more about the unknown user's preferences to improve its knowledge (exploration). This problem has been addressed by the reinforcement learning community but they do not consider the risk level of the current user's situation, where it may be dangerous to explore the non-top-ranked documents the user may not desire in his/her current situation if the risk level is high. 
We introduce in this paper an algorithm named CBIR-R-greedy that considers the risk level of the user's situation
to adaptively balance between exr and exp.
\end{abstract}

\section{Introduction}
\label{intro}
A considerable amount of research has been done in information retrieval interesting content for mobile users. Earlier techniques in Context-Based Information Retrieval (CBIR) \cite{Goenka2010, Rani2013, 24} are based solely on the computational behaviour of the user to model his interests regarding his surrounding environment like location, time and near people (the user's situation). The main limitation of such approaches is that they do not take into account the dynamicity of the user's content. 
This gives rise to another category of IR techniques, like \cite{Hofmann2011} that tries to tackle this limitation by using on-line learning.
Such systems obtain feedback on the few top-ranked documents results and also from other documents, by exploring the non-top-ranked documents that could lead to a better solution. However, the system also needs to ensure that the quality of result lists is high by exploiting what is already known. Clearly, this results in an exr/exp dilemma studied in the “Bandit algorithm” \cite{21}.
The challenge for the existent contextual bandit algorithms is to construct result lists from several documents, so that one result list contains both exploratory and exploitative documents and the algorithms have to choose the number of each of them in that list.
We introduce in this paper an algorithm, named CBIR-R-greedy, that computes the optimal value of exploration according to the risk of the user situation. 
We consider as risky or critical, a situation where it is dangerous to explore the non-top-ranked documents; this means that it is not desired, can yield to a trouble, or causes a waste of time for the user when reading a document which is not interesting for him in the current situation. 
In this case, the exploration-oriented learning 
should be avoided. 
CBIR-R-greedy extends the CBIR-$\epsilon$-greedy \cite{BouneffoufBG13a} strategy with an update of exr/exp by selecting suitable user's situations for either exr or exp. We have tested CBIR-R-greedy in an on-line evaluation with professional mobile users. 
The rest of the paper is organized as follows. Section 2 reviews some related works. Section 3 describes the algorithms involved in the proposed approach. The experimental evaluation is illustrated in Section 4. Finally, Section 5 concludes the paper and points out possible directions for future work.
\section{Related works}
\label{sec:2}
We refer, in the following to a state of the art on CBIR and also techniques that tackle both making dynamic exr/exp (bandit algorithm) and considering the risk in ranking the result. 

\textbf{Result ranking in CBIR Result ranking in CBIR.} Different recent works address the challenge of ranking IR results according to the user's context.
In \cite{Rani2013}, a linear function of reordering is adopted to adjust the search results to the user. The initial score of the document is multiplied by a score of customization, which is a linear combination as a weighted sum of the preferences of user content and location. 
To rank documents in \cite{Rani2013}, a pre rank is calculated to measure how much a document is related to a particular context and the post rank is measured to know how much a document is related to user query. 
The total rank weight is calculated by summing up pre rank weight and post rank weight. Documents are ranked in descending order of their final rank weight.
In \cite{Goenka2010} the contextual user's profile is used to reorder the search results. They propose a linear combination of the original result returned by a conventional IR with a score of customization calculated between the document and the profile of the user's document ranking. 
The personalization score is calculated for each document with the contextual user's profile based on a similarity measure between the vector of weighted concepts of the document and the user profile is also represented as a vector of weighted concepts.
\\
As shown above, none of the mentioned works address the exr/exp problem in CBIR.

\textbf{Bandit Algorithms for CBIR.} The multi-armed bandit problem was originally described by Robbins \cite{21}.
Compared to the standard multi-armed bandit problem, the CBIR does not select individual documents, but constructs result lists from several documents, so that one result list contains both exploratory and exploitative documents. Therefore, the bandit algorithms need to be modified to manage this new challenge. Authors in \cite{Hofmann2011} have studied the contextual bandit problem in IR area. They have proposed to adapt the $\epsilon$-greedy approach to their need. They maintain two document lists, one exploitative (based on the currently learned best ranking), and one exploratory (introducing variations to the current best ranking to explore potential improvements). An exploration rate $\epsilon$ determines the relative number of documents each list contributes to the final list shown to the user. However this rate is just left to the user and is not deeply studied.
A recent work done by \citep{BouneffoufBG13a} proposed to also adapt the $\epsilon$-greedy approach to CBIR, where they consider some situations as critical and they propose to manage the exploration trade-off according to that situations.

\textbf{The Risk-Aware Decision.}
The risk-aware decision has been studied for a long time in reinforcement learning, where the risk is defined as the reward criteria that not only takes into account the expected reward, but also some additional statistics of the total reward, such as its variance or standard deviation \cite{Luenberger1998InvestmentScience}.
In RS the risk is recently studied. The authors in \cite{BouneffoufBG12, bouneffouf2013drars, } consider the risk of the situations in the recommendation process, and the study yields to the conclusion that considering the risk level of the
situation on the exr/exp strategy significantly increases the performance of the recommender system.

\textbf{Discussion.} 
As shown above, their is only one work that addresses the exp/exp in CBIR; where they just consider the similarity between situations to get the risk of the current situation. 

We propose to improve this work by improving their risk computing as follows:

(1) The risk is computed by aggregating three approaches including the approach proposed in \cite{BouneffoufBG13a} :

Handling semantic concepts to express situations and their associated risk level, the first approach compute
the risk using concepts. This approach permits to get the risk of the situation directly from the risk of each of its concepts. The second approach compute the risk using the semantic similarity between the current situation and situations stocked in the system as it is done in \cite{BouneffoufBG13a}, and it comes from the assumption that similar situations have the same risk level. The third approach is computing the risk using the variance of the reward. In this case, we assume that risky situations get very low number of user's clicks. 
 (2) We propose an algorithm called CBIR-R-greedy that include the risk computing in its management of the exr/exp trade-off. High exploration (resp. high exploitation) is achieved when the current user situation is "not risky" (resp. "risky"); 
\section{The Proposed CBIR Model} 
\label{sec:crs}
This section focuses on the proposed model and starts by introducing the key notions used in this paper.
\\
\textbf{Situation:} A situation is an external semantic interpretation of low-level context data, enabling a higher-level specification of human behaviour. More formally, a situation $S$ is a n-dimensional vector $S=(O_{\delta_{1}}.c_1,O_{\delta_{2}}.c_2,...,O_{\delta_{n}}.c_n)$ where each $c_i$ is a concept of an ontology $O_{\delta_{i}}$ representing a context data dimension. 
According to our need, we consider a situation as a 3-dimensional vector 
$S=(O_{Location}.c_i, O_{Time}.c_j, O_{Social}.c_k)$ where $c_i, c_j, c_k $ are concepts of Location, Time and Social ontologies.
\\ 
\textbf{ \textit{User's interests}:} The user's interests are represented by using the most representative terms derived from the assumed relevant documents in a particular search situation. In particular, let $q_i$ be the query submitted by a specific user to the retrieval situation $S^i$. We assume that a document retrieved by the search engine with respect to $q_i$ is relevant if it causes a user's click. Let $D^i$ be the set of assumed relevant documents in situation $S^i$. Then, $UI^i$ (the user's interests) corresponds to the vector of weighted terms in $D^i$, where the weight $w_{tm}$ of term $tm$ is computed as follows:  $w_{tm} = \frac{1}{|D^i|} \sum_{d \in D^t}tf(tm,d)*log(n/n_{tm})$, where $tf(tm; d)$ is the frequency of term $tm$ in document $d\in D^i$, $n$ is the number of documents in the collection, $n_{tm}$ is the number of documents in the collection containing $tm$. Each document $d \in D^i$ is represented by a term vector where the relevance value of each term $tm$ in situation $S^i$ is computed using the $tf * idf$ weighting.
\\
\textbf{ \textit{The user model}:} The user model is structured as a case base composed of a set of situations with their corresponding $UI$, denoted $UM=\{(S^i; UI^i)\}$, where $S^i \in S$ is the user situation and
$UI^i \in UI $ its corresponding user interests.
\\
\textbf{\textit{Definition of risk:}} "The risk in information retrieval is the possibility to disturb or to upset the user (which leads to a bad answer of the user)".
\\
From the precedent definition of the risk, we have proposed to consider in our system Critical Situations (CS) which is a set of situations where the user needs the best information that can be retrieved by the system, because he can not be disturbed. This is the case, for instance, of a professional meeting. In such a situation, the system must exclusively perform exploitation rather than exploration-oriented learning. In other cases where the risk of the situation is less important (like for example when the user is using his information system at home, or he is on holiday with friends), the system can make some exploration by retrieving information without taking into account his interest.
\\
To consider the risk level of the situation in RS, we go further in the definition of situation by adding it a risk level $R$, as well as one to each concept: 
$S [R]$=($O_{\delta_{1}}.c_1 [cv_1]$ $,O_{\delta_{2}}.c_2 [cv_2],...,O_{\delta_{n}}.c_n [cv_n]$) where 
$CV$=$\{cv_1, cv_2,...,cv_n\}$ is the set of risk levels assigned to concepts, $cv_ i \in [0,1]$. $R\in [0,1]$ is the risk level of situation $\textit{S}$, and the set of situations with $R=1$ are considered as critical situations (CS). 

We propose CBIR to be modelled as a contextual bandit problem including user's situation information.  
Formally, a bandit algorithm proceeds in discrete trials $t=1...T$. For each trial $t$, the algorithm performs the following tasks:
\\
\textbf{Task 1:} Let $S^t$ be the current user's situation when he/she submits a request, and 
$PS$ the set of past situations. The system compares $S^t$ with the situations in \textit{PS} in order to choose the most similar one, $S^p = argmax_{S^{i}\in PS}sim(S^t,S^i)$. The semantic similarity metric is computed by:
$sim(S^t,S^i)=\frac{1}{|\Delta|}\sum_{\delta \in \Delta}sim_{\delta}(c^t_{\delta},c^i_{\delta})$, where $sim_{\delta}$ is the similarity metric related to dimension $\delta$ between two concepts $c_{\delta}^t$ and $c_{\delta}^i$, and $\Delta$ is the set of dimensions (in our case Location, Time and Social).
The similarity between two concepts of a dimension $\delta$ depends on how closely $c_{\delta}^t$ and $c_{\delta}^i$ are related in the corresponding ontology. To compute $sim_{\delta}$, we use the same similarity measure as \cite{15}:
$sim_{\delta}(c_{\delta}^{t},c_{\delta}^{i})=2*\frac{deph(LCS)}{deph(c_{\delta}^t)+ deph(c_{\delta}^i)}$, where $LCS$ is the Least Common Subsumer of $c_{\delta}^t$ and $c_{\delta}^i$, and $deph$ is the number of nodes in the path from the current node to the ontology root. 
\\
\textbf{Task 2:} Let $D$ be the documents collection. After retrieving $S^p$, the system observes the corresponding user’s interests $UI^p$ in the user's model case base. Based on the observed $UI^p$ and the query $q$, the algorithm ranks documents in $D$ using the traditional cosine similarity measure.
\\
\textbf{Task 3:} From the ranked list of documents presented to the user, the algorithm receives the set $D^t$ of clicked documents and improves its document-selection strategy with the new observation: in situation $S^t$, documents in $D^t$ obtain a user's click. Depending on the similarity between the current situation $S^t$ and its most similar situation $S^p$, two scenarios are possible:
\\
(1) If sim($S^t, S^p$) $<$ 1: the current situation does not exist in the case base; the system adds this new case composed of the current situation $S^t$ and the current user's interest $UI^t$ computed from the set $D^t$ of clicked documents using Eq. 1; 
\\
(2) If sim($S^t, S^p$) = 1: the situation exists in the case base; the system updates the case having as premise the situation $S^p$ with the current user's interest $UI^t$, the update being done by integrating the new documents, $D^p = D^p \bigcup D^c$, and computing the new vector $UI^t$.

\textbf{The IR-$\epsilon$-greedy algorithm}
The IR-$\epsilon$-greedy algorithm ranks documents using the following equation:
   \begin{equation}
   % \label{eq:4}
   \label{alg:ucb}
   %\scriptsize{} 
   d_t =\left\{ \begin{array}{rcl}
   		result_q(q,d_i) & if( l<j; q < \epsilon) \\ 
                  result_c(UI^p,d_i)  & if( l\geq j; l < \epsilon) \\
                   Random(d_i)  & if( l \geq \epsilon) 
             \end{array}\right.\\ 
   \end{equation}             
In Eq. 5, $i\in{1,…N}$ where $N$ is the number of documents in the collection, $d_i \in D^t$, $l$ and $j$ are random values uniformly distributed over [0, 1] which define the exr/exp trade-off; $\epsilon \in[0, 1]$ is the probability of making a random exploratory rank of documents. $result_q(q,d_i)$ gives the original score returned by the system based on the query $q$ using the cosine similarity as follows:  $result_q(q,d_i) = \frac {q.d_i} {||q|| ||d_i||}$            
,where $result_c(UI^p,d^i)$ gives the contextualization score returned by the system based on the user's interests $UI^p$, and it is also computed using cosine similarity as follows : $result_c(UI^p,d_i) = \frac{UI^p.d_i}{||UI^p|| ||d_i||}$. Random($d^i$) gives a random rank to the document $d_i$ to perform exploration.  

\textbf{The CBIR-R-greedy Algorithm.} To improve the adaptation of the IR-$\epsilon$-greedy algorithm to the risk level of the situations, the CBIR-R-greedy algorithm (Alg. \ref{alg:rucb}) computes the probability of exploration $\epsilon$ by using the situation risk level $R(S^t)$ as indicated in subsection \ref{sec:Risk}.     
\begin{algorithm}[H]
\caption{The CBIR-R-greedy algorithm}
%   \label{alg:2} 
\label{alg:rucb} 
\begin{algorithmic}[1]
\STATE {\bfseries Input:} $S^t, D^p, RD = \emptyset, B, N, \epsilon _{min}, \epsilon_{max}$  
\STATE {\bfseries Output:} $RD$ 
\STATE $\epsilon = \epsilon _{max}-R(S^t) \times (1-\epsilon _{min}) $                  
\STATE $RD$ = IR-$\epsilon$-greedy($\epsilon, D^p, RD, N$)       
   \end{algorithmic}
\end{algorithm}

\subsection{Computing the Risk Level of the Situation}
\label{sec:Risk}
The risk complete level \textit{$R(S^t)$} of the current situation is computed by aggregating three approaches $R^c$, $R^v$ and $R^m$ as follows:
\begin{equation}
\label{eq:rst}
R(S^t)=\sum_{j \in J}\lambda_{j}R_{j}(S^t)
%\vspace*{- 0.1 cm}     
\end{equation}  
In Eq.~\ref{eq:rst}, $R_{j}$ is the risk metric related to dimension $j \in J$, where $J=\{m, c, v\}$; $\lambda_{j}$ is the weight associated to dimension $j$ and it is set out using an off-line evaluation. $R_c$ compute the risk using concepts, $R_m$ compute the risk using the semantic similarity between the current situation and situations stocked in the system and $R_v$ compute the risk using the variance of the reward. In what follows, we describe the three approaches and their aggregation.

\textbf{Risk Computed using the Variance of the Reward.}
To compute the risk of the situation using the variance of the reward, we suppose that the distribution of the Click Through Rate (the CTR is the number of clicks per recommendation) of the situations follows a normal distribution. From this assumption, and according to confidence interval theory \cite{Wald1942}, we compute the risk using Eq.~\ref{eq:Rv}. Here, the idea is that, more the CTR of situations is low (low number of user's clicks) more the situation is risky. 
\begin{equation}
 \label{eq:Rv}
 % R_v(S^t) = E(CTR(S)) - \alpha*\frac{\sigma(CTR(S))}{\sqrt{|S|}}\\ 
  R_v(S^p) = \left\{ \begin{array}{rcl}
  	1- \frac{CTR(S^p) - Var}{1-Var}& \mbox{if} & CTR(S^p) > Var\\ 
       1  &   &   \mbox{Otherwise}
            \end{array}\right.\\ 
\end{equation}

In Eq.~\ref{eq:Rv}, the risk threshold $Var= E(CTR(S)) - \alpha*\sigma(CTR(S))$, where $\sigma$ is the variance of $CTR(S)$ and $\alpha$ is constant fixed to 2 according to Gauss theory \cite{Wald1942}. The $CTR(S)= \frac{click(S)}{rec(S)}$, where $click(S)$ gives the number of times that the user clicks in documents recommended in $S$ and $rec(S)$ gives the number of times that the system has made recommendation in the situation $S$. 

\textbf{Risk Computed using Concepts.}
Computing the risk using concepts gives a weighted mean of the risk level of the situation concepts:
%%\vspace*{- 0.3 cm} 
\begin{equation}
\label{eq:Rc}
R_c(S^t) = \sum_{\delta \in \Delta} \mu_{\delta} cv_{\delta}^{t} \qquad \mbox{if}\qquad CV \neq \emptyset    
\end{equation}
In Eq.~\ref{eq:Rc}, $cv_{\delta}^{t}$ is the risk level of dimension $\delta$ in $S^t$ and $\mu_{\delta}$ is the weight associated to dimension $\delta$, set out by using an arithmetic mean $\mu_{\delta}=\frac{1}{|CS|}(\sum_{S^i \in CS} cv_{\delta}^{i})$, the idea behind that is to make the mean of all the risk levels associated to concepts related to the dimension $\delta$ in $CS$.
 
\textbf{Risk Computed using Semantic Similarity between the Current Situation and Past Situations.}
The risk may also be computed using the semantic similarity between the current situation and $CS$ stocked in the system. This permits to give the risk of the situation from the assumption that a situation is risky if it is similar to a pre-defined $CS$.
\\ 
The risk $R_m(S^t)$ is computed using Eq.~\ref{eq:Rm} 
\begin{equation}
% \label{eq:4}
\label{eq:Rm}
%\scriptsize{} 
R_m(S^t) =\left\{ \begin{array}{rcl}
		 1- B + sim(S^t,S^m)  & if \hspace{.1cm} sim(S^t,S^m) < B \\ 
              \hspace{-.5cm}  1  & otherwise 
          \end{array}\right.\\ 
\end{equation} 
In Eq.~\ref{eq:Rm}, the risk is extracted from the degree of similarity between the
current situation $S^t$ and the centroid critical situation $S^m$. $B$ is the similarity threshold and it is computed using an off-line simulation. From Eq.~\ref{eq:Rm}, we see that the situation risk level $R_m(S^t)$ increases when the similarity between $S^t$ and $S^m$ increases.
The critical situation centroid is selected from $CS$ as follows: $S^m=argmax_{S^{f}\in{CS}}\frac{1}{|CS|}\sum_{S^e\in{CS}}sim(S^f,S^e)$.

\textbf{Updating the Risk Value.}
\label{sec:4}
After the retrieval process and the user's feedback, the system propagates the risk to the concepts of the ontology using Eq.~\ref{eq:cv} and propagates the risk in $CS$ using Eq.~\ref{eq:CSS} :   
 \begin{equation}
 \label{eq:cv}
\forall cv \in S^t \;\; cv = \frac{1}{|CV_{cv}|}(\sum_{S^{i} \in CV_{cv}} cv_{i}^{\eta})\\
 \end{equation}
The idea in Eq.~\ref{eq:cv} is to make the mean of all the risk levels associated to concepts $cv$ related to situations $S^{i}$ in the user's situation historic for the dimension $\eta$. In Eq.~\ref{eq:cv}, $CV_{cv}$ gives the set of situations where $cv$ has been computed.
 \begin{equation}
 \label{eq:CSS}
R(S^t) = \frac{1}{T}(\sum_{k=1}^{k=T} R(S^{t}_{k}))\\
 \end{equation}
The idea in Eq.~\ref{eq:CSS} is to make the mean of all the risk levels associated to the situation $S^t$ in the user's situation historic. In Eq.~\ref{eq:CSS}, $k \in [0,T] $ gives the number of times that the risk of $S^t$ is computed.
   
\section{Experimental Evaluation}
\label{sec:experimental}
In order to empirically evaluate the performance of our approach, and in the absence of a standard evaluation framework, we propose an online evaluation framework. The main objective of the experimental evaluation is to evaluate the performance of the proposed algorithm (CBIR-R-greedy). In the following, we present and discuss the obtained results. We have conducted a diary study with the collaboration of a French software company which provides an enterprise search engine application that allows users to connect their information system through mobile devices. 
\\
We conduct our experiment with 3500 users. We compare CBIR-R-greedy to the state of the art algorithm CBIR-$\epsilon$-greedy used in \cite{BouneffoufBG13a} we called it here CBIR-Rm-greedy and also the IR-$\epsilon$-greedy with exploration $\epsilon$=0 (non exploration algorithm, baseline). We have done that to verify the correlation between commuting the risk of the situation and the performance of the algorithm.
To this end, we have randomly split users on three groups: the first group has an IR system with the CBIR-R-greedy; the second group is equipped with CBIR-Rm-greedy; finally, the last group uses the IR-$\epsilon$-greedy with exploration $\epsilon$=0 (non exploration algorithm, baseline). 
\\
Note that we do not evaluate the algorithm proposed in \cite{Hofmann2011} because it does not consider the variance of the $\epsilon$, which is our goal in this evaluation. 
Two experimental evaluations have been carried out, as follows.

\textbf{Precision on the top 10 documents} 
We compare the algorithms regarding precision of the system on the top 10 documents, which is the number of clicks on the top 10 documents per the number of times the users make a request. In Fig. 1, the horizontal axis represents the day of the month and the vertical axis is the performance metric.
\begin{center}
  \begin{figure}%[htb]
  % DANS CETTE FIGURE, CE N'EST PAS ITERATIONS BUT SIZE OF DATA
  \begin{center}
  \includegraphics[width=0.9\textwidth]{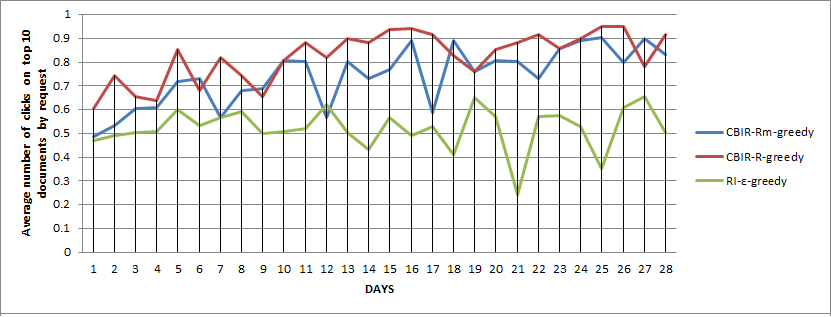}\\
  \caption{Average precision on top 10 documents for exr/exp algorithms}  
  \label{fig:9} 
 \end{center}  
  \end{figure}   % For one-column wide figures use  
\end{center}   
From Fig. 1 we observe  that overall, tested algorithms have better performances than the baseline. While the CBIR-Rm-greedy algorithm converges to a higher precision compared with a baseline, its overall performance is not as good as CBIR-R-greedy. 
\\
We have also observed the average number of clicks per request for all the 28 days and we observe that the CBIR-R-greedy algorithm effectively has the best average precision during this month, which is 0.82 . CBIR-Rm-greedy obtains  a precision of 0.74 and the baseline, 0.52 . CBIR-R-greedy increases the average precision by a factor of 1.54 over the baseline and a factor of 1.1 over the CBIR-Rm-greedy algorithm. The improvement comes from a dynamic trade-off between exr/exp on documents ranking, controlled by the risk estimation. We observe that better the risk level of the situation is computed better is the result.

\textbf{Average time spent on documents} 
We look at the time spent on documents to evaluate the relevance of the clicked documents. Fig. 2 gives the average time spent in each document by day. 
\begin{center}
  \begin{figure}%[htb]
  % DANS CETTE FIGURE, CE N'EST PAS ITERATIONS BUT SIZE OF DATA
  \begin{center}
  \includegraphics[width=0.9\textwidth]{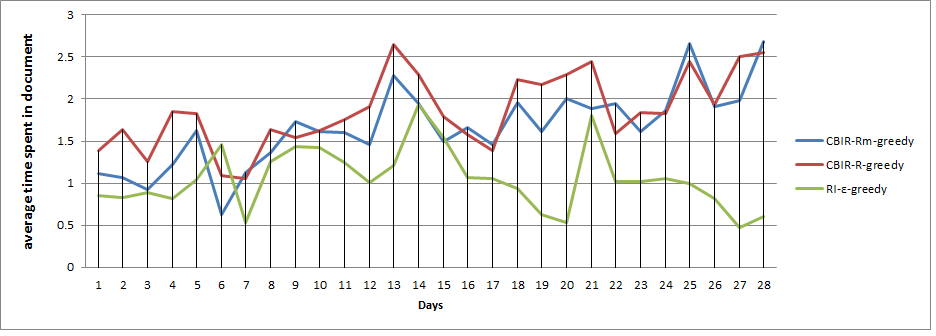}\\
  \caption{Average time spent on document for exr/exp algorithms}  
  \label{fig:9} 
 \end{center}  
  \end{figure}   % For one-column wide figures use  
\end{center}  
We have computed the average time spent per document for all the 28 days and we observe that the CBIR-R-greedy algorithm effectively has the best average number of time spent in documents during this month, which is 1.86 minutes.
CBIR-Rm-greedy obtains 1.65 minutes and the baseline, 1.06 minutes. CBIR-R-greedy increases the average time spent by a factor of 1.12 over the CBIR-Rm-greedy. We can conclude that the CBIR-R-greedy increases the relevance of clicked documents by it best strategy in computing the risk of the situation. 
             	
\section{Conclusion}
In this paper, we have studied the problem of exploitation and exploration in Context-Based Information Retrieval and proposed a novel approach that ranks documents by balancing adaptively exr/exp regarding the risk level of the situation.
We have presented an on-line evaluation protocol with real mobile user. We have evaluated our approach according to the proposed evaluation protocol. This study leads to the conclusion that considering the risk the exr/exp trade-off, significantly increases the performance of the CBIR.
In the future, we plan to improve the notion of situation level risk in the scope of CBIR by introducing more contextual dimensions. 

\bibliographystyle{abbrv} %ieeetr  abbrv
\bibliography{example_paper} 

\end{document}